\documentclass[twoside,twocolumn,english,superscriptaddress,nofootinbib,preprintnumbers, aps]{revtex4}

\usepackage{euscript,amssymb}
\usepackage{amsthm}
\usepackage{graphicx}
\usepackage{amsfonts}
\usepackage{amsmath}
\usepackage{amssymb}
\usepackage{fancyhdr}
\usepackage{epsfig}
\usepackage{color}
\usepackage{esint}
\usepackage[unicode=true, pdfusetitle, bookmarks=true,bookmarksnumbered=false,bookmarksopen=false, breaklinks=false,pdfborder={0 0 1},backref=false,colorlinks=false]{hyperref}

\begin{document}

%%%%%%%%%%%%%%%%%%%%%%%%%%%%%%%%%%%%%%%%%%%%%%%%%%%%%%%%%%%%%%%%%%%%%%%%%%%%%%%%%%%%%
\title{Lorentz Symmetry in QFT on Quantum Bianchi I  Space-Time}
%%%%%%%%%%%%%%%%%%%%%%%%%%%%%%%%%%%%%%%%%%%%%%%%%%%%%%%%%%%%%%%%%%%%%%%%%%%%%%%%%%%%%

\author{Andrea Dapor}
\email{adapor@fuw.edu.pl} \affiliation{Instytut Fizyki Teoretycznej, Uniwersytet Warszawski, ul. Ho\.{z}a 69, 00-681 Warsaw, Poland}

\author{Jerzy Lewandowski}
\email{jerzy.lewandowski@fuw.edu.pl} \affiliation{Instytut Fizyki Teoretycznej, Uniwersytet Warszawski, ul. Ho\.{z}a 69, 00-681 Warsaw, Poland}

\author{Yaser Tavakoli}
\email{tavakoli@ubi.pt} \affiliation{Departamento de F\'{\i}sica,
Universidade da Beira Interior, Rua Marqu\^{e}s d'Avila e Bolama, 6200 Covilh\~{a}, Portugal}

%%%%%%%%%%%%%%%%%%%%%%%%%%%%%%%%%%%%%%%%%%%%%%%%%%%%%%%%%%%%%%%%%%%%%%%%%%%%%%%%%%%%%%%%%%%
\begin{abstract}

We develop the quantum theory of a scalar field on LQC Bianchi I geometry. In particular, by considering only the single modes of the field, the evolution equation is derived from the quantum scalar constraint; it is shown that the same equation can be obtained from QFT on a ``classical'' effective geometry. We then study the dependence of this effective space-time on the wave-vector of the modes (which could in principle generate a deformation in local Lorentz symmetry), by analyzing the dispersion relation for propagation of the test field on the resulting geometry. We show that when we disregard the back-reaction no Lorentz-violation is present, despite the effective metric being different than the classical Bianchi I one. Furthermore, a preliminary analysis of the correction due to inclusion of back-reaction is briefly discussed in the context of Born-Oppenheimer approximation.

\end{abstract}

\date{\today}

\pacs{04.60.-m, 04.60.Pp, 98.80.Qc}

\maketitle

\section{Introduction}
\label{Intro}

Quantum field theory (QFT) in curved space-time is a
theory wherein matter is treated quantum-mechanically, but gravity is treated
classically in agreement with general relativity \cite{mw,Birrell}. Despite its classical treatment of gravity,
QFT on curved space-time has provided a good approximate description in circumstances
where the quantum effects of gravity do not play a dominant role.
On the other hand, during the cosmological era arbitrarily close to the classical singularity, such effects cannot be neglected and the theory is no longer valid.
This suggests that the background classical space-time in the Planck regime near the
singularity has to be replaced by a quantum background \cite{Fredenhagen}.

Loop quantum cosmology (LQC) \cite{LQC-ref1,LQC-ref2} has become in the last few years an interesting candidate for a quantum description
of the early universe at the Planck scale, providing a number of concrete results.
This approach, that comprises a family of symmetry-reduced cosmological models within the framework of Loop Quantum Gravity (LQG) \cite{alrev,crbook,ttbook},
was first fully applied to the spatially $k = 0$ FRW cosmology coupled to a massless scalar
field (that also serves as an internal time parameter \cite{aps1,aps2,aps3}). It was shown numerically that the
big bang singularity is replaced by a quantum bounce, and that the energy density is bounded by a critical density $\rho_{\text{cr}}\sim 0.41 \rho_{\text{Pl}}$ of the order of the Planck density (see also \cite{mb1,abl,acs,cs1,kl}).
Furthermore, extensions of this quantization method were successfully exposed for $k=0$ models
with or without cosmological constant \cite{bp,ap}, $k = 1$ closed models \cite{apsv,warsaw}, $k = -1$ open models \cite{kv}, and
for the simplest anisotropic cosmology (namely, Bianchi I \cite{awe2}, Bianchi II \cite{awe3}, and Bianchi IX \cite{awe4} models).

Discrete approaches to quantum gravity lead to a breakdown of the usual structure of space-time at around the Planck scale,
with possible violations of Lorentz symmetry.
This can have phenomenological implications, such as a deformation of the dispersion
relations for propagating particles (modes of a matter field) \cite{ACamelia1,ACamelia2,ACamelia3}. These modifications would change the speed of propagation,
introducing delays/advances for particles of different energies. These can be in principle detected in experimental tests such as observation of gamma ray bursts \cite{ACamelia1}.

Of course, to study such effects in the framework of LQC, one needs to couple a quantum field to a LQC space-time: in other words, to develop QFT on LQC geometry.
From a field-theoretic perspective, LQC provides a family of well-defined quantum geometries on which
quantum matter can propagate. The first step in the construction of QFT on such quantum geometries has been developed in \cite{AKL},
for a real massless scalar field on $k = 0$ FRW quantum space-time. Therein, a mode decomposition of the quantum fields was given,
so that the field could be regarded as a collection of decoupled harmonic oscillators. This allows one to study each quantum mode
separately, by ignoring the infinite number of degrees of freedom of the field.
In that work, it was shown that each mode of the field ``sees'' the
underlying quantum geometry as an effective ``classical'' one. In principle, such effective geometry could depend on the wave-vector of the mode, thereby producing
an (apparent) Lorentz-violation: different quanta would simply propagate on different space-times. It turns out that in the FRW case this effective metric is the same for all modes,
and thus no Lorentz-violation is present.
A possible explanation of this lies in the fact that FRW space-time is conformally flat, and therefore massless fields (such as the one
used in \cite{AKL}) do not distinguish it from Minkowski space-time. If this is the case, then a more significant test would be to consider more complicate space-times, such as Bianchi type I model.

The purpose of this paper is to generalize the model of QFT on FRW cosmological space-time to the one of the Bianchi I model,
and analyze the issue of Lorentz symmetry in this context.
In light of the Belinskii-Khalatnikov-Lifshitz (BKL) conjecture \cite{BKL1,BKL2}, such space-times are the most interesting ones, since they are more likely to represent the cosmological dynamics in the Planck scale regime when the matter source (without any symmetry assumption) is a massless scalar field \cite{BKL3}.
On the other hand, as already said above, it is hoped that anisotropy would lead to a wave-vector-dependent effective geometry, thus producing a modification in the particles' dispersion relation.

The paper is organized as follows. We start in section \ref{BImodel} by briefly reviewing the LQC-like quantization of Bianchi I geometry.
In section \ref{QFT-CG-QG}, we consider a test scalar field on this background Bianchi I space-time:
in subsection \ref{QFT-CG} we summarize the essential features of QFT on a classical Bianchi I background geometry;
in subsection \ref{QFT-QG} we develop QFT on quantum Bianchi I space-time.
In section \ref{BI-Effective} the dynamics of QFT on classical and quantum backgrounds are compared, and an effective ``classical'' geometry is shown to emerge.
In section \ref{Phenomenology} we investigate the possibility that quanta of field of different energies could in fact probe different effective geometries, i.e. different
aspects of the underlying quantum geometry, possibly producing Lorentz-violation. In particular, we compare the propagation of the test field on classical and effective geometries, derive the modified dispersion relation and study the status of causality.
In section \ref{conclusions} we discuss the results and conclusions. Appendix \ref{ap-modes} complements our discussion
on mode decompositions of the test field, while appendix \ref{app2} probes the consistency of our findings under a different choice of physical time.

%%%%%%%%%%%%%%%%%%%%%%%%%%%%%%%%%%%%%%%%%%%%%%%%%%%%%%%%%%%%%%%%%%%%%%%%
\section{Background Quantum Geometry}
\label{BImodel}
%%%%%%%%%%%%%%%%%%%%%%%%%%%%%%%%%%%%%%%%%%%%%%%%%%%%%%%%%%%%%%%%%%%%%%%%%

In this section we present a brief summary of LQC of Bianchi I space-time.

%%%%%%%%%%%%%%%%%%%%%%%%%%%%%%%%%%%%%%%%%%%%%%%%%%%%%%%%%%%%%%%
\subsection{Classical Bianchi I space-time}
\label{BImodel-1}

Let us consider the background space-time manifold $M$ to be topologically $M = \mathbb{R} \times \mathbb{T}^{3}$, where $\mathbb{T}^{3}$ is the 3-torus (whose coordinates $x^{i}$ range in $(0, 1)$). As is standard in the literature on
Bianchi models, we will restrict ourselves to the diagonal Bianchi I metric:
\begin{eqnarray} \label{1}
g_{\mu \nu} dx^{\mu} dx^{\nu} = -dt^{2} + \sum_{i = 1}^{3} a_{i}^{2}(t) (dx^{i})^{2}.
\label{metric1}
\end{eqnarray}
This choice of coordinates is not the most general: we have chosen the lapse function for the time coordinate $t$ as $N_{t} = 1$, and the shift vector $N_{t}^{a} = 0$. In LQG the canonical pair consists of the gravitational $SU(2)$ connection $A^{a}_{i}$ and the densitized triad $E^{i}_{a}$, which for metric (\ref{metric1}) are given by
\begin{eqnarray} \label{3}
A^{i}_{a} = c^{i} \delta^{i}_{a}, \ \ \ \ \ E^{a}_{i} = p_{i} \delta^{a}_{i},
\end{eqnarray}
where $c_i$, $p^i$ are constants. Therefore, the phase space of gravity, $\Gamma_{\text{gr}}$, is coordinatised by the canonical variables $(c^{i}, p_{i})$, satisfying the canonical Poisson algebra
\begin{eqnarray}
\{c^{i}, p_{j}\} = 8 \pi G \gamma \delta^{i}_{j}\ ,
\label{Poisson}
\end{eqnarray}
with $\gamma$ the Immirzi parameter. The relations between the phase space variables $p_i$ and the scale factors $a_{i}$ are
\begin{eqnarray}
p_{1} = a_{2} a_{3}, \ \ \ \ \ p_{2} = a_{3} a_{1}, \ \ \ \ \ p_{3} = a_{1} a_{2}.
\label{5}
\end{eqnarray}

The matter source for this model will be a massless scalar field $T$, which will also serve as the relational time.
Therefore, it is convenient to work with a harmonic time function $\tau$, i.e., satisfying $\square\tau = 0$.
Since $\tau$ is homogeneous, the spatial part of this equation vanishes, and hence we get $d^{2}\tau/dt^{2} = 0$. Integrating this over the fiducial cell $\mathcal{V}$, we get
\begin{eqnarray}
\int d^{4}x \frac{d^{2}\tau}{dt^{2}} = \int dt \frac{d^{2}\tau}{dt^{2}} \int_{\mathcal{V}} d^{3}x = \frac{d\tau}{dt} V=\text{const}.,
\label{9}
\end{eqnarray}
where $V$ is the volume of the fiducial cell $\mathcal{V}$, which is given in terms of the gravitational variables as
\begin{eqnarray}
V = |a_{1} a_{2} a_{3}| = \sqrt{|p_{1} p_{2} p_{3}|}\ .
\label{10}
\end{eqnarray}
For a choice of the integration constant being unit, we then get $dt/d\tau = V$. This object allows us to change the space-time coordinates from $(t, x^{i})$ to $(\tau, x^{i})$: the corresponding lapse function, $N_\tau$, can be obtained using the relation $N_{\tau} d\tau = N_{t} dt$:
\begin{eqnarray}
N_{\tau} = N_{t} \frac{dt}{d\tau} = \sqrt{|p_{1} p_{2} p_{3}|}\ .
\label{11}
\end{eqnarray}
In terms of the new coordinates $(\tau, x^{i})$, the Bianchi I metric (\ref{metric1}) becomes then
\begin{align}
g_{\mu \nu} dx^{\mu} dx^{\nu} & =  -N_{\tau}^{2} d\tau^{2} + \sum_{i = 1}^{3} a_{i}^{2}(\tau) (dx^{i})^{2}
\notag \\
& =  |p_{1} p_{2} p_{3}| \left[-d\tau^{2} + \sum_{i = 1}^{3} \frac{(dx^{i})^{2}}{p_{i}^{2}} \right].
\label{12}
\end{align}

General relativity is a constrained theory, and in Ashtekar variables (\ref{3}) the constraints comprise so-called Gauss constraint, vector constraint and scalar constraint. However, since we have restricted ourselves to diagonal homogeneous metrics and fixed the internal gauge,
the Gauss and the vector constraints are identically satisfied. The only non-trivial constraint (i.e., the homogeneous part of the scalar constraint) can be expressed by restricting the integration in the full theory to the fiducial cell ${\cal V}$:
\begin{align}
C_{\text{gr}} & =  \int_{\mathcal{V}} d^{3}x\ N_{\tau} {\cal C}_{\text{gr}} = -\frac{1}{8 \pi G \gamma^{2}} \int_{\mathcal{V}} d^{3}x  \frac{N_{\tau}}{|p_{1} p_{2} p_{3}|}  \notag \\
&\ \ \ \ \ \times (p_{1} p_{2} c_{1} c_{2} + p_{2} p_{3} c_{2} c_{3} + p_{3} p_{1} c_{3} c_{1})  \notag \\
& =  -\frac{1}{8 \pi G \gamma^{2}} (p_{1} p_{2} c_{1} c_{2} + p_{2} p_{3} c_{2} c_{3} + p_{3} p_{1} c_{3} c_{1}).
\label{13}
\end{align}
On the other hand, the massless scalar field $T$ and its conjugate momentum $P_{T}$ coordinatise the phase space of matter, denoted by $\Gamma_{T}$. The energy density for the massless scalar field is given by $\rho_{T} = P_{T}^{2}/2V^{2}$, so the contribution of $T$ to the scalar constraint can be obtained as
\begin{eqnarray}
C_{T} = \int_{\mathcal{V}} d^{3}x N_{\tau} {\cal C}_{T} = \frac{P_{T}^{2}}{2}\ ,
\label{14}
\end{eqnarray}
where we used ${\cal C}_{T}=\sqrt{q}\rho_{T}$.

Putting together the two contributions (\ref{13}) and (\ref{14}),
we obtain the total scalar constraint for the system (namely, the scalar constraint of the geometry, denoted by $C_{\text{geo}}$) as
\begin{eqnarray}
C_{\text{geo}} = C_{\text{gr}} + C_{T}.
\label{15}
\end{eqnarray}
Physical states of the classical geometry lie on the constraint surface defined on $\Gamma_{\text{geo}} = \Gamma_{T} \times \Gamma_{\text{gr}}$ by the equation $C_{\text{geo}} = 0$. The $\tau$-evolution of any phase space function $f$ is obtained by the Poisson brackets:
\begin{eqnarray}
\frac{df}{d\tau} = \{f, C_{\text{geo}}\}.
\label{16}
\end{eqnarray}
In particular, the $\tau$-evolutions of the scalar field $T$ and and its conjugate momentum $P_{T}$ read,
\begin{align}
\dfrac{dT}{d\tau} &= \{T, C_{\text{geo}}\}
= P_{T}, \label{17-a} \\
\dfrac{dP_{T}}{d\tau} &= \{P_{T}, C_{\text{geo}}\} = 0.
\label{17}
\end{align}
From equation (\ref{17}) it is seen that the conjugate momentum $P_{T}$ of the scalar field $T$ is a constant of motion (which is chosen here to be positive), and hence equation (\ref{17-a}) describes a linear evolution of the scalar field $T$ with respect to $\tau$:
\begin{eqnarray}
T\ =\  P_{T} \tau\ ,
\label{18}
\end{eqnarray}
for a vanishing integration constant. This equation implies that the phase space variable $T$ is a good time parameter for $\tau$-evolution of the system. In other words, although $T$ does not have the physical dimensions of time, it is a good evolution parameter; in LQC literature it is known as the \emph{relational time}. Using the relation $N_T dT=N_\tau d\tau$ in (\ref{17-a}), we get an expression for the lapse function $N_T$:
\begin{eqnarray}
N_{T}\ = \ \frac{\sqrt{|p_{1} p_{2} p_{3}|}}{P_{T}} \ .
\label{19}
\end{eqnarray}
Thus, we can change coordinates from $(\tau, x^{i})$ to the physical $(T, x^{i})$ one. Then, the metric (\ref{12}) becomes
\begin{eqnarray}
g_{\mu \nu} dx^{\mu} dx^{\nu} = |p_{1} p_{2} p_{3}| \left[-\frac{1}{P_{T}^{2}} dT^{2} + \sum_{i = 1}^{3} \frac{(dx^{i})^{2}}{p_{i}^{2}} \right].
\label{20}
\end{eqnarray}

%%%%%%%%%%%%%%%%%%%%%%%%%%%%%%%%%%%%%%%%%%%%%%%%%%%%%%%%%%%%%%%%%%
\subsection{Quantum Bianchi I space-time}
\label{BImodel-2}

The previous construction was preparatory for loop quantization, which we sketch now. The kinematical Hilbert space of Bianchi I model, ${\cal H}_{\text{kin}}$, is given as the tensor product of the Hilbert space $\mathcal{H}_{\text{gr}}$ of the gravitational sector and the Hilbert space $\mathcal{H}_{\text{T}}$. More precisely, the states in $\mathcal{H}_{\text{gr}}$ can be labeled by 3 real numbers, which are arranged in a vector $\vec{\lambda}$ \cite{awe2}.
The gravitational part of $C_{\text{geo}}$ is turned into a difference operator acting on $\mathcal{H}_{\text{gr}}$.

On the other hand, the scalar field sector is quantized according to Schroedinger picture: the Hilbert space is $\mathcal{H}_{\text{T}} = L_{2}(\mathbb{R}, dT)$, while the dynamical variables $T$ and $P_{T}$ are promoted to operators on it:
\begin{eqnarray} \label{21}
\widehat{T} = T, \ \ \ \ \ \widehat{P}_{T} = -i\hbar \frac{\partial}{\partial T} \ .
\end{eqnarray}
Therefore,  the kinematical Hilbert space of geometry is given by $\mathcal{H}_{\text{kin}} = \mathcal{H}_{\text{T}} \otimes \mathcal{H}_{\text{gr}}$. The scalar constraint operator $\widehat{C}_{\text{geo}}$ is well-defined on $\mathcal{H}_{\text{kin}}$ and is given by
\begin{eqnarray}
\widehat{C}_{\text{geo}} = -\frac{1}{2} (\hbar^{2}\partial_{T}^{2} \otimes \mathbb{I}) - \frac{1}{2} (\mathbb{I} \otimes \Theta)\ .
\label{22}
\end{eqnarray}
where $\Theta$ is a difference operator on $\mathcal{H}_{\text{gr}}$ defined in \cite{awe2}.

Physical states of the quantum geometry are those $\Psi_o(T, \vec{\lambda}) \in \mathcal{H}_{\text{kin}}$ lying in the kernel of $\widehat{C}_{\text{geo}}$. In other words, they are solutions to equation
\begin{eqnarray}
-\hbar^{2} \partial_{T}^{2} \Psi_o(T, \vec{\lambda}) = \Theta \Psi_o(T, \vec{\lambda})\ .
\label{23}
\end{eqnarray}
One can further restrict solutions of the equation (\ref{23}) to the space spanned by the ``positive frequency'' solutions, i.e., the solutions to
\begin{align}
-i\hbar \partial_{T} \Psi_o(T, \vec{\lambda}) & = \sqrt{|\Theta|} \Psi_o(T, \vec{\lambda}) \notag \\
& =: \widehat{H}_o\Psi_o(T, \vec{\lambda}) \  .
\label{23-a}
\end{align}
We will take this as our physical Hilbert space of geometry, $\mathcal{H}_{\text{phys}}^o$, endowed with scalar product
\begin{eqnarray}
\langle \Psi_{o} | \Psi'_{o} \rangle = \sum_{\lambda_{1}, \lambda_{2}, \lambda_{3}}  \overline{\Psi_{o}(T_o, \vec{\lambda})} \Psi'_{o}(T_o, \vec{\lambda}) ,
\label{23-c}
\end{eqnarray}
where $T_o$ is any ``instant" of internal time $T$.

%%%%%%%%%%%%%%%%%%%%%%%%%%%%%%%%%%%%%%%%%%%%%%%%%%%%%%%%%%%%%%%%%%%%%%%%%%%%%%%%%%%
\section{The Test Quantum Field}
\label{QFT-CG-QG}
%%%%%%%%%%%%%%%%%%%%%%%%%%%%%%%%%%%%%%%%%%%%%%%%%%%%%%%%%%%%%%%%%%%%%%%%%%%%%%%%%%%

This section is divided into two parts. In the first, for the convenience of the reader, we present a brief summary of the essential
features of  QFT on a classical Bianchi I background space-time. In the second,
we consider the case of a quantum Bianchi I background, as presented in the previous section.

%%%%%%%%%%%%%%%%%%%%%%%%%%%%%%%%%%%%%%%%%%%%%%%%%%%%%%%%%%%%%%%%%%%
\subsection{QFT on classical Bianchi I background}
\label{QFT-CG}

We consider the background space-time to be the classical Bianchi I model (of section \ref{BImodel-1}), equipped with the coordinates $(x_0, x^j)$, in which $x^j\in \mathbb{T}^3$,
with $x_0\in \mathbb{R}$ being a generic time coordinate. In this coordinates, the Bianchi I background space-time is described by the metric
\begin{eqnarray}
g_{\mu \nu} dx^{\mu} dx^{\nu} = -N_{x_{0}}^{2}(x_{0}) dx_{0}^{2} + \sum_{i = 1}^{3} a_{i}^{2}(x_{0}) (dx^{i})^{2}.
\label{24}
\end{eqnarray}
Let us consider a real (inhomogeneous) test scalar field\footnote{This is to be considered a \emph{test field},
so that it does not induce any back-reaction on the geometry, which can then be thought of as \emph{background} geometry.} $\phi(x_0,\vec{x})$ on this background space-time, whose Lagrangian is
\begin{eqnarray}
\mathcal{L}_{\phi} = \frac{1}{2} (g^{\mu \nu} \partial_{\mu} \phi \partial_{\nu} \phi - m^{2} \phi^{2}).
\label{26}
\end{eqnarray}

Performing the Legendre transformation, one gets the canonically conjugate momentum for the test field $\phi$, denoted by $\pi_\phi$, on a $x_{0} = \text{const}$ slice.
Then, for the pair $(\phi,\pi_\phi)$, the classical solutions of the equation of motion (coming from (\ref{26})) can be expanded in Fourier modes:
\begin{align}
& \phi(x_{0}, \vec{x}) = \frac{1}{(2\pi)^{3/2}} \sum_{\vec{k} \in \mathcal{L}} \phi_{\vec{k}}(x_{0}) e^{i \vec{k} \cdot \vec{x}}, \notag \\
& \pi_{\phi}(x_{0}, \vec{x}) = \frac{1}{(2\pi)^{3/2}} \sum_{\vec{k} \in \mathcal{L}} \pi_{\vec{k}}(x_{0}) e^{i \vec{k} \cdot \vec{x}},
\label{27}
\end{align}
where $(k_{1}, k_{2}, k_{3}) \in (2\pi \mathbb{Z})^{3}$ span a 3-dimensional lattice $\mathcal{L}$ (see appendix \ref{ap-modes}).

Using the Fourier transform (\ref{27}) in Poisson algebra of $\phi$ and $\pi_{\phi}$, it follows that the Fourier coefficients $\phi_{\vec{k}}$ and $\pi_{\vec{k}}$ satisfy
$\{\phi_{\vec{k}}, \pi_{\vec{k}'}\} = \delta_{\vec{k}, -\vec{k}'}$ and the reality conditions
$\phi_{\vec{k}} = \overline{\phi_{-\vec{k}}}$ and $\pi_{\vec{k}} = \overline{\pi_{-\vec{k}}}$.
The Hamiltonian follows from (\ref{26}):
\begin{align}
H_{\phi}(x_{0}) & =  \frac{1}{2} \int d^{3}x \frac{N_{x_{0}}}{\sqrt{|p_{1} p_{2} p_{3}|}} \left[\pi_{\phi}^{2} + \sum_{i = 1}^{3} (p_{i} \partial_{i} \phi)^{2} \right.  \notag \\
& \ \ \ \left. + |p_{1} p_{2} p_{3}| m^{2} \phi^{2}\right] \notag \\
& =  \frac{N_{x_{0}}}{2\sqrt{|p_{1} p_{2} p_{3}|}} \sum_{\vec{k} \in \mathcal{L}}  \left[\overline{\pi_{\vec{k}}} \pi_{\vec{k}} \right. \notag \\
& \ \ \ \left. + \left(\sum_{i = 1}^{3} (p_{i} k_{i})^{2} + |p_{1} p_{2} p_{3}| m^{2}\right) \overline{\phi_{\vec{k}}} \phi_{\vec{k}}\right].
\label{30}
\end{align}

In appendix \ref{ap-modes} it is shown that, by a wise choice of variables, it is possible to rewrite (\ref{30}) as the Hamiltonian for a collection of decoupled harmonic oscillators. Specifically we obtain:
\begin{align}
& H_{\phi}(x_{0})\  := \ \sum_{\vec{k} \in \mathcal{L}} H_{\vec{k}}(x_{0}) = \frac{N_{x_{0}}}{2 \sqrt{|p_{1} p_{2} p_{3}|}} \notag \\
&\ \ \   \times \sum_{\vec{k} \in \mathcal{L}} \left[p_{\vec{k}}^{2} + \left(\sum_{i = 1}^{3} (p_{i} k_{i})^{2} + |p_{1} p_{2} p_{3}| m^{2}\right) q_{\vec{k}}^{2}\right] ,
\label{37-a}
\end{align}
where $q_{\vec{k}}$ and $p_{\vec{k}}$ are the two conjugate variables associated with the $\vec{k}$ modes.
Notice that, for each mode $\vec{k}$, the term $H_{\vec{k}}(x_0)$ in equation (\ref{37-a}) is nothing but the Hamiltonian of a single harmonic oscillator with a time-dependent mass/frequency.

In this paper we will focus on the dispersion relation  of the quantum field (see section \ref{Phenomenology}), for which studying a single mode $q_{\vec{k}}$ is sufficient. On the other hand, the quantization of the full system would require to take into account all modes. In particular, renormalisation of the UV limit is a crucial element of the definition of the QFT: without it, expressions such as equation (\ref{37-a}) are entirely formal. Considering such an infinite-dimensional system is not at all straightforward, and leads to the whole topic of QFT in curved space-time. However, as long as one is interested only in a single mode (or a finite set), quantization is on the line of quantum harmonic oscillator: the Hilbert space is $\mathcal{H}_{\vec{k}} = L_{2}(\mathbb{R}, dq_{\vec{k}})$, and the dynamical variables $q_{\vec{k}}$ and $p_{\vec{k}}$ are promoted to operators on it, $\widehat{q}_{\vec{k}}\psi(q_{\vec{k}}) = q_{\vec{k}}\psi(q_{\vec{k}})$ and
$\widehat{p}_{\vec{k}}\psi(q_{\vec{k}}) = -i\hbar \partial/\partial q_{\vec{k}}\psi(q_{\vec{k}})$.
Time evolution (with respect to generic $x_{0}$) is generated by the time-dependent Hamiltonian operator $\widehat{H}_{\vec{k}}(x_{0})$ via Schroedinger equation:
\begin{align}
& i\hbar \partial_{x_{0}} \psi(x_{0},q_{\vec{k}})  =  \widehat{H}_{\vec{k}}(x_{0}) \psi(x_{0}, q_{\vec{k}}) \notag \\
&\ \ \ \ \ =  \frac{N_{x_{0}}(x_{0})}{2 \sqrt{|p_{1}(x_{0}) p_{2}(x_{0}) p_{3}(x_{0})|}} \left[\widehat{p}_{\vec{k}}^2 + \left(\sum_{i = 1}^{3} (p_{i} k_{i})^{2} \right. \right. \notag \\
& \ \ \ \ \ \ \ \ \left. \left. + |p_{1} p_{2} p_{3}| m^{2}\right) \widehat{q}_{\vec{k}}^{2}\right] \psi(x_{0}, q_{\vec{k}}).
\label{44}
\end{align}
This concludes the study of QFT on classical Bianchi I background. However, to consider QFT on quantum Bianchi I background, it is necessary to fix the choice of time in accordance with the previous section. In the $x_{0} = \tau$ case, it is $N_{\tau} = \sqrt{|p_{1} p_{2} p_{3}|}$, and (\ref{44}) reduces to
\begin{align}
i\hbar \partial_{\tau} \psi(\tau,q_{\vec{k}}) & =  \widehat{H}_{\tau,\vec{k}}(\tau) \psi(\tau, q_{\vec{k}}) \notag \\
& =  \frac{1}{2} \left[\widehat{p}_{\vec{k}}^2 % \notag \\
+ \omega_{\tau,\vec{k}}^2 \widehat{q}_{\vec{k}}^{2}\right] \psi(x_{0}, q_{\vec{k}}),
\label{44-b}
\end{align}
where we defined the (time-dependent) frequency
\begin{align}
\omega_{\tau, \vec{k}}^2 := \sum_{i = 1}^{3} (p_{i} k_{i})^{2} + |p_{1} p_{2} p_{3}| m^{2}.
\label{omeeega}
\end{align}
for any $\vec{k}$ mode.

%%%%%%%%%%%%%%%%%%%%%%%%%%%%%%%%%%%%%%%%%%%%%%%%%%%%%%%%%%%%%%%%%
\subsection{QFT on quantum Bianchi I background}
\label{QFT-QG}

We consider a system of general relativity coupled to an homogeneous massless scalar field $T$, together with a massive and (in general) inhomogeneous test field $\phi$, propagating on the homogeneous background. The scalar constraint on the full phase space of the gravitational field, scalar field $T$ and test field $\phi$, is expressed as
\begin{equation}
C = C_{\text{gr}} + C_{T} + C_{\phi}.
\label{total-class}
\end{equation}
Let us fix the time coordinate $\tau$ with the lapse function $N_{\tau} = \sqrt{|p_1p_2p_3|}$ for the whole system. Then, the scalar constraint becomes
\begin{equation}
C_{\tau} = C_{\text{geo}}(\tau) + H_\phi(\tau),
\label{totalH-Class}
\end{equation}
where $C_{\text{geo}}$ is given in equations (\ref{15}) and $H_\phi(\tau)$ is obtained from (\ref{37-a}):
\begin{eqnarray}
H_\phi(\tau) = \sum_{\vec{k} \in \mathcal{L}} H_{\tau, \vec{k}}=\frac{1}{2}\sum_{\vec{k} \in \mathcal{L}}\left[p_{\vec{k}}^{2} + \omega_{\tau, \vec{k}}^{2} q_{\vec{k}}^{2}\right].
\label{46}
\end{eqnarray}

To pass to the quantum theory, let us focus just on a single mode $\vec{k}\in \mathcal{L}$. The total kinematical Hilbert space for the system is given by the tensor product ${\cal H}_{\text{kin}}^{(\vec{k})} = {\cal H}_{\text{geo}} \otimes L^2(\mathbb{R}, dq_{\vec{k}})$.
Therefore, from (\ref{totalH-Class}) the scalar constraint for the QFT (of a single mode $\vec{k}$) on the quantum Bianchi I background is given by
\begin{align}
\widehat{C}_{\tau, \vec{k}}\ & :=\  \widehat{C}_{\text{geo}} + \widehat{H}_{\tau, \vec{k}} = -\frac{\hbar^{2}}{2} (\partial_{T}^{2} \otimes \mathbb{I}_{\text{gr}} \otimes \mathbb{I}_{\vec{k}}) \notag \\
&\ \ \ - \frac{1}{2}(\mathbb{I}_{T} \otimes \Theta \otimes \mathbb{I}_{\vec{k}}) + (\mathbb{I}_{T} \otimes \widehat{H}_{\tau, \vec{k}}).
\label{48-a}
\end{align}
The Hamiltonian operator $\widehat{H}_{\tau, \vec{k}}$ in equation (\ref{48-a}) is given by
\begin{eqnarray}
\widehat{H}_{\tau, \vec{k}} = \frac{1}{2} \left[\widehat{p}_{\vec{k}}^{2} + \left(\sum_{i = 1}^{3} \widehat{p}_{i}^{2} k_{i}^{2} + |\widehat{p}_{1} \widehat{p}_{2} \widehat{p}_{3}| m^{2}\right) \widehat{q}_{\vec{k}}^{2}\right],
\label{47}
\end{eqnarray}
acting on the kinematical Hilbert space $\mathcal{H}_{\text{gr}} \otimes \mathcal{H}_{\vec{k}}$. Notice that, although $\widehat{p}_{i}$ are operators, they commute with each other and with matter operators, $\widehat{p}_{\vec{k}}$ and $\widehat{q}_{\vec{k}}$, and so does the ``frequency operator'' (in round brackets).

Physical states, $\Psi(T,\vec{\lambda}, q_{\vec{k}})$, are those $\Psi \in \mathcal{H}^{(\vec{k})}_{\text{kin}}$ that satisfy
\begin{eqnarray}
-\hbar^{2} \partial_{T}^{2} \Psi(T, \vec{\lambda}, q_{\vec{k}}) = \left[\widehat{H}_{o}^{2} - 2 \widehat{H}_{\tau, \vec{k}}\right] \Psi(T, \vec{\lambda}, q_{\vec{k}}),
\label{49}
\end{eqnarray}
where $\widehat{H}_{o} = \sqrt{|\Theta|}$ as above. Since the operator $\widehat{H}_{o}^{2} - 2 \widehat{H}_{\tau, \vec{k}}$ on the right hand side of (\ref{49}) is symmetric on ${\cal H}_{\text{kin}}^{(\vec{k})}$,  we can restrict to its self-adjoint extension on a suitable domain\footnote{Here we are assuming that such extension exists. In case this is not true for the operator $\Theta$, one should seek an alternative quantization, such that $\widehat{H}_{o}^{2} - 2 \widehat{H}_{\tau, \vec{k}}$ admits a self-adjoint extension. In fact, what follows does not rely on the details of operator $\widehat{H}_{o}$.}.
On the physical Hilbert space, this operator is identified with $\widehat{P}^2_{T}$, which is classically a positive Dirac observable; thus, we can consider just the positive part of the spectrum of $\widehat{H}_{o}^{2} - 2 \widehat{H}_{\tau, \vec{k}}$.
\begin{proof}[\textbf{Remark}]
Consider the $m = 0$ isotropic case for definiteness. Classically, for large volumes $v \sim p^{3/2}$ it is $H_o^2 \sim v^2$ and $H_{\tau, \vec{k}} \sim v^{4/3} k^2$, so the classical inequality $H_o^2 - 2 H_{\tau, \vec{k}} \geq 0$ can always be satisfied for $k \gg 1$ provided that $v$ is large enough. Quantum-mechanically, one uses the boundedness of operator $\Theta$ by $\rho_{\text{cr}} v^2$ (a property independent of the matter content). From Einstein equation it follows that $\widehat{H}_{\tau, \vec{k}} = \widehat{H}_o^2 - \widehat{P}_T^2$, and hence $\widehat{H}_{\tau, \vec{k}}$ itself is bounded in a similar way. Since $\widehat{H}_{\tau, \vec{k}}$ is the Hamiltonian of an harmonic oscillator with frequency proportional to $v^{2/3} k$, the ground state of this operator has eigenvalue proportional to $v^{2/3} k/2$. Because of boundedness, such frequency must be less then or equal to $\rho_{\text{cr}} v^2$ for all $k$. This sets a $k$-dependent lower bound to the volume,
\begin{eqnarray} \label{minvol}
v \geq \left(\frac{k}{2 \rho_{\text{cr}}}\right)^{3/4} =: v_{min},
\end{eqnarray}
which semiclassical states of geometry achieve at the bounce. We thus see that $k$ can be arbitrarily large: the geometry will ``react'' by concentrating on large volumes.
\end{proof}
From these considerations, it follows that the subspace of the kinematical Hilbert space where $\widehat{H}_o^2 - 2 \widehat{H}_{\tau, \vec{k}}$ is positive-definite is non-empty for all $\vec{k}$: restricting to such a subspace, we are able to define the square root of operator $\widehat{H}_{o}^{2} - 2 \widehat{H}_{\tau, \vec{k}}$. Therefore, physical states are solutions to the Schroedinger-like equation
\begin{align}
-i\hbar \partial_{T} \Psi(T, \vec{\lambda}, q_{\vec{k}}) & = \left[\widehat{H}_{o}^{2} - 2 \widehat{H}_{\tau, \vec{k}}\right]^{1/2}\Psi(T, \vec{\lambda}, q_{\vec{k}}).
\label{50}
\end{align}
The space spanned by these solutions is the physical Hilbert space $\mathcal{H}^{(\vec{k})}_{\text{phys}}$ of the theory (for the chosen mode $\vec{k}$). The scalar product on the physical Hilbert space is simply
\begin{eqnarray}
\langle \Psi | \Psi' \rangle = \sum_{\lambda_{1}, \lambda_{2}, \lambda_{3}} \int_{-\infty}^{\infty} dq_{\vec{k}} \overline{\Psi(T_{o}, \vec{\lambda}, q_{\vec{k}})} \Psi'(T_{o}, \vec{\lambda}, q_{\vec{k}}).
\label{51}
\end{eqnarray}

If the test field approximation holds, the back-reaction of the field $\phi$ is neglected on the homogeneous background geometry. Therefore, the theory is physically relevant only when we treat $2\widehat{H}_{\tau, \vec{k}}$ as a perturbation to $\widehat{H}_o^2$. Using this assumption, we obtain from (\ref{50}) the following approximation (for a discussion, see \cite{AKL})
\begin{align}
-i \hbar \partial_{T} \Psi(T, \vec{\lambda}, q_{\vec{k}}) &=  \left[\widehat{H}_{o} - \widehat{H}_{o}^{-\frac{1}{2}} \widehat{H}_{\tau, \vec{k}} \widehat{H}_{o}^{-\frac{1}{2}}\right] \Psi(T, \vec{\lambda}, q_{\vec{k}}) \notag \\
&=:  \left[\widehat{H}_{o} - \widehat{H}_{T, \vec{k}}\right] \Psi(T, \vec{\lambda}, q_{\vec{k}}).
\label{53}
\end{align}
In the last step we introduced the operator $\widehat{H}_{T, \vec{k}}$ which implements the Hamiltonian that generates evolution in relational time $T$. More precisely, at the classical level, the Hamiltonian that generates evolution in $T$ can be found by considering the lapse function $N_T=P_T^{-1}N_\tau$ in equation (\ref{37-a}): it is $H_{T, \vec{k}} = H_{\tau, \vec{k}} N_{T}/N_{\tau} = H_{\tau, \vec{k}} P_{T}^{-1} \approx H_{\tau, \vec{k}} H_{o}^{-1} $ (the last step holding thanks to the test field approximation, which at physical level allows us to put $P_{T} = (H_{o}^{2} - 2 H_{\tau, \vec{k}})^{1/2} \approx H_{o}$). At the quantum level, the operator $\widehat{H}_{T, \vec{k}}$ in equation (\ref{53}) is a particular quantum realization of the classical function $H_{T, \vec{k}}$.

%%%%%%%%%%%%%%%%%%%%%%%%%%%%%%%%%%%%%%%%%%%%%%%%%%%%%%%%%%%%%%%%%%%%%%%%
\section{Effective Bianchi I Geometry}
\label{BI-Effective}
%%%%%%%%%%%%%%%%%%%%%%%%%%%%%%%%%%%%%%%%%%%%%%%%%%%%%%%%%%%%%%%%%%%%%%%%

In the previous section, we have constructed QFT on a Bianchi I quantum background geometry, obtaining equation (\ref{53}) as the quantum counterpart of equation (\ref{44-b}) for QFT on classical space-time.
In this section, by a comparison between these two dynamical equations, we will discuss the relation between the aspects of QFT in classical
and in quantum geometry. To do this, we shall now construct a classical geometry limit of the QFT on quantum Bianchi I space-time.

At the quantum geometry level, equation (\ref{53}) provides a quantum evolution equation for the state $\Psi(T,\vec{\lambda},q_{\vec{k}})$, depending on (the $\vec{k}$-th mode of) the test field $\phi$ \emph{and} the quantum geometry (encoded in $\vec{\lambda}$). On the other hand, on the classical geometry, $\psi(T,q_{\vec{k}})$ given in equation (\ref{44-b}), is the evolution of the state of the test field $\phi$ on the time-dependent classical background geometry.
To compare the two frameworks, we shall consider the \emph{test field approximation}, that is, we disregard the back-reaction of matter on geometry. Thus, we can write the state of the system as a tensor product of the state of geometry and the state of matter, as the two are disentangled:
\begin{align}
\Psi(T, \vec{\lambda}, q_{\vec{k}}) = \Psi_o(T, \vec{\lambda}) \otimes \psi(T, q_{\vec{k}}),
\label{zerothstate}
\end{align}
where the geometry evolves through $\widehat{H}_o$, i.e., $-i\hbar \partial_T \Psi_o = \widehat{H}_o \Psi_o$ (see equation (\ref{23-a})); in other words, it is $\Psi_o(T, \vec{\lambda}) = e^{iT\widehat{H}_o/\hbar} \Psi_o(0, \vec{\lambda})$. Plugging this in equation (\ref{53}) and projecting both sides on $\Psi_o(T, \vec{\lambda})$, one finds an equation for the matter only:
\begin{align}
& i \hbar \partial_{T} \psi(T, q_{\vec{k}}) = \frac{1}{2} \left[\langle \widehat{H}_{o}^{-1} \rangle\widehat{p}_{\vec{k}}^{2}
+ \langle \widehat{H}_{o}^{-\frac{1}{2}} \left(\sum_{i = 1}^{3} \widehat{p}_{i}^2(T) k_{i}^{2} \right. \right. \notag \\
&\ \  \  \ \ \left. \left.  + |\widehat{p}_{1}(T) \widehat{p}_{2}(T) \widehat{p}_{3}(T)| m^{2}\right) \widehat{H}_{o}^{-\frac{1}{2}} \rangle \widehat{q}_{\vec{k}}^{2}\right] \psi(T, q_{\vec{k}}),
\label{58}
\end{align}
where we moved the $T$-dependence from $\Psi_o$ to the gravitational operators (i.e., we describe the geometry sector in Heisenberg picture); so that, $\langle \widehat{A}(T) \rangle$ denotes the expectation value on the quantum state of geometry $\Psi_{o}(0, \vec{\lambda})$ of gravitational operator
\begin{align}
\widehat{A}(T) = e^{-iT \widehat{H}_o/\hbar} \widehat{A} e^{iT \widehat{H}_o/\hbar} \ .
\label{heispicture}
\end{align}

Equation (\ref{58}) is an evolution equation for (the $\vec{k}$-th mode of) a quantum field on an ``effective" geometry. Let us assume that this geometry is described by a metric $\bar{g}_{\mu \nu}$ of the form
\begin{align}
\bar{g}_{\mu \nu} dx^{\mu} dx^{\nu} = -\bar{N}^{2}(T) dT^{2} + |\bar{p}_{1} \bar{p}_{2} \bar{p}_{3}| \sum_{i = 1}^{3} \frac{(dx^{i})^{2}}{\bar{p}_{i}^{2}} \ .
\label{effMetr}
\end{align}
Then, by setting $x_0 = T$ in equation (\ref{44}) we obtain
\begin{align}
i\hbar \partial_{T} \psi(T,q_{\vec{k}}) & = \widehat{H}_{\vec{k}}(T) \psi(T, q_{\vec{k}}) \notag \\
& = \frac{\bar{N}_{T}(T)}{2 \sqrt{|\bar{p}_{1} \bar{p}_{2} \bar{p}_{3}|}} \left[\widehat{p}_{\vec{k}}^2
+ \left(\sum_{i = 1}^{3} (\bar{p}_{i} k_{i})^{2} \right. \right. \notag \\
& \ \ \ \left. \left. + |\bar{p}_{1} \bar{p}_{2} \bar{p}_{3}| m^{2}\right) \widehat{q}_{\vec{k}}^{2}\right] \psi(T, q_{\vec{k}}),
\label{44-c}
\end{align}
which is a classical time-dependent evolution equation for (the $\vec{k}$-th mode of) the test field with respect to $T$. Now, by comparing (\ref{44-c}) and (\ref{58}), we find the following relations:
\begin{align}
\bar{N}(T) = \langle \widehat{H}_{o}^{-1} \rangle \sqrt{|\bar{p}_{1} \bar{p}_{2} \bar{p}_{3}|} ,
\label{60-a}
\end{align}
\begin{align}
\dfrac{\bar{N}(T)}{\sqrt{|\bar{p}_{1} \bar{p}_{2} \bar{p}_{3}|}} \bar{p}_{i}^{2} =  \langle \widehat{H}_{o}^{-\frac{1}{2}} \widehat{p}_{i}^{2}(T) \widehat{H}_{o}^{-\frac{1}{2}} \rangle ,
\label{60-b}
\end{align}
\begin{align}
\bar{N}(T)  m^{2}  = m^{2} \frac{\langle \widehat{H}_{o}^{-\frac{1}{2}} |\widehat{p}_{1}(T) \widehat{p}_{2}(T) \widehat{p}_{3}(T)| \widehat{H}_{o}^{-\frac{1}{2}} \rangle}{ \sqrt{|\bar{p}_{1} \bar{p}_{2} \bar{p}_{3}|}} \ .
\label{60-c}
\end{align}
These relations hold only if we ignore the quantum fluctuations of the geometry, and assume that the underlying quantum geometry state $\Psi_{o}$ is peaked on an effective trajectory (with expectation values $\bar{P}_T$ and $\bar{p}_i$ of $\widehat{H}_o$ and $\widehat{p}_i$, respectively). In cosmological applications, the semiclassical states $\Psi_{o}$ of geometry have very small dispersions along the entire effective trajectory, thus, these assumptions are justified.

Let us consider now the system of equations (\ref{60-a}), (\ref{60-b}) and (\ref{60-c}). There are five equations with four variables, $\bar{N}_{T}$ and $\bar{p}_{i}$, and hence, there exists no general solutions\footnote{This of course does not mean that massive scalar fields cannot exist on quantum geometry. It simply suggests that there is no effective geometry of the form (\ref{effMetr}) that reproduces the equation of motion of such a field on the quantum geometry. Moreover, note that (\ref{60-a}) and (\ref{60-c}) are classically equivalent, so we can expect that what follows also applies to the massive case, as long as we disregard quantum fluctuations (that is to say, we consider only semiclassical states of geometry). Nevertheless, for definiteness we will set $m = 0$ in the following.}.
However, considering a massless test field $\phi$, the last equation is identically verified, and the remaining equations reduce to a solvable system of four equations with four variables. It turns out that there exists a unique solution for $\bar{N}_{T}$ and $\bar{p}_{i}$ given by:
\begin{align}
& \bar{N}(T) = \langle \widehat{H}_{o}^{-1} \rangle^{1/4} \left(\prod_{i = 1}^{3} \langle \widehat{H}_{o}^{-\frac{1}{2}} \widehat{p}_{i}^{2}(T) \widehat{H}_{o}^{-\frac{1}{2}} \rangle\right)^{\frac{1}{4}}, \label{61-a} \\
& \bar{p}_{i} = \left[\dfrac{\langle \widehat{H}_{o}^{-1/2} \widehat{p}_{i}^{2}(T) \widehat{H}_{o}^{-1/2} \rangle}{\langle \widehat{H}_{o}^{-1} \rangle}\right]^{\frac{1}{2}}.
\label{61-b}
\end{align}
These relations determine the effective space-time (\ref{effMetr}) in terms of expectation values of  the gravitational operators on
the quantum geometry state $\Psi_{o}$. It should be noticed that the effective metric components (\ref{61-a}) and (\ref{61-b}) do not depend on $\vec{k}$ which indicates that all modes of the field ``probe" the same effective background geometry, or their energy, independently. For this reason we expect that no Lorentz-violation is present, and indeed it can be shown that there is no quantum gravity effect on the dispersion relation for the field $\phi$: we will study this issue in the next section.

To conclude, it is worth mentioning that the effective geometry defined in (\ref{61-a}) and (\ref{61-b}) comes from the $m = 0$ choice, or more precisely the $m \ll \|\vec{k}\|$ limit. In this limit, equation (\ref{60-c}) can be disregarded, and one finds the presented result. On the other hand, one may consider the opposite limit, namely $m \gg \|\vec{k}\|$. In this case, the three equations in (\ref{60-b}) are to be disregarded, and the system is underdeterminate. However, if one assumes that both $m$ and $\|\vec{k}\|$ are large, then it is equation (\ref{60-a}) that drops from the system, and one has again a unique solution:
\begin{align}
& \bar{N}_T = \left(\frac{\prod_i \langle \widehat{H}_o^{-\frac{1}{2}} \widehat{p}_i^2 \widehat{H}_o^{-\frac{1}{2}} \rangle}{\langle \widehat{H}_o^{-\frac{1}{2}} | \widehat{p}_1 \widehat{p}_2 \widehat{p}_3 | \widehat{H}_o^{-\frac{1}{2}} \rangle}\right)^{\frac{1}{2}},
\label{alt61-a} \\
& \bar{p}_i = \langle \widehat{H}_o^{-\frac{1}{2}} | \widehat{p}_1 \widehat{p}_2 \widehat{p}_3 | \widehat{H}_o^{-\frac{1}{2}} \rangle \left(\frac{\langle \widehat{H}_o^{-\frac{1}{2}} \widehat{p}_i^2 \widehat{H}_o^{-\frac{1}{2}} \rangle}{\prod_j \langle \widehat{H}_o^{-\frac{1}{2}} \widehat{p}_j^2 \widehat{H}_o^{-\frac{1}{2}} \rangle}\right)^{\frac{1}{2}}.
\label{alt61-b}
\end{align}
As already pointed out in the footnote above, the effective metric defined in (\ref{alt61-a}) and (\ref{alt61-b}) is classically equivalent to the one defined in (\ref{61-a}) and (\ref{61-b}). Nevertheless, if fluctuations of the semiclassical state of geometry could consistently be taken into account, it would seem that two different behaviours are to be expected from light and heavy particles respectively.

%%%%%%%%%%%%%%%%%%%%%%%%%%%%%%%%%%%%%%%%%%%%%%%%%%%%%%%%%%%%%%%%%%%%%%%%%%%%%%%%%%%%%%%%%%%%%
\section{Dispersion Relations and Lorentz Symmetry}
\label{Phenomenology}

A prediction of many approaches to quantum gravity comes from the study of \emph{in vacuo} ``dispersion relation".
This is the relation between the frequency $\omega$ and the wave-vector $\vec{k}$ of a mode of a
field \emph{in vacuo}. In QFT, the dispersion relation links the momentum and the energy of quanta, and thus for fundamental fields its form is dictated by Lorenz symmetry (namely, by the mass-shell constraint).

Deformed dispersion relations are a rather natural possibility in quantum gravity \cite{ACamelia1,ACamelia2,ACamelia3}, requiring a modification of Lorentz symmetry, which is then said to be ``broken'' by quantum gravity effects. If that is the case, Lorentz-invariance is only an approximate symmetry of the low-energy world.
Hence, quantum gravity effects could be observed by relying
on modified dispersion relations for the matter (such as photons or neutrinos) traveling on a
quantum gravitational background. More precisely, Planck scale effects are expected to be negligible
in standard circumstances, but observations of highly energetic particles traveling long
distances may set bounds on violation of (standard) Lorentz symmetr. Some of these properties may be experimentally detectable in satellite facilities (e.g. GLAST or AMS), using as probes light from distant astrophysical sources, such as gamma ray bursts (GRBs).

We analyze the form of the effective geometry obtained in the previous section, addressing the question on whether a Lorentz-violation is produced in our model. To do this, let us first consider the Hamiltonian of the test field on the classical background given by equation (\ref{37-a}). This equation determines the evolution of modes $q_{\vec{k}}$ in time coordinate $x_{0}$. More precisely, for each mode $\vec{k}\in {\cal L}$, the  $x_0$-evolution of each pair of variables $(q_{\vec{k}}, p_{\vec{k}})$ is given by Hamilton equations:
\begin{align}
\dfrac{dq_{\vec{k}}}{dx_{0}} = \{q_{\vec{k}}, H_{\vec{k}}\}, \ \ \ \dfrac{dp_{\vec{k}}}{dx_{0}} = \{p_{\vec{k}}, H_{\vec{k}}\},
\label{38-a}
\end{align}
where $H_{\vec{k}}$ is defined by equation (\ref{H-k}). Using equation (\ref{38-a}) one can get the wave equation for each mode $q_{\vec{k}}$ (see appendix \ref{ap-modes}). For the case $x_0=\tau$, the lapse function is $N_{\tau} = \sqrt{|p_{1} p_{2} p_{3}|}$, and we get the wave equation for each $q_{\vec{k}}$ (see equation (\ref{42})) as
\begin{eqnarray}
\frac{d^{2}q_{\vec{k}}}{d\tau^{2}} + \omega_{\tau, \vec{k}}^{2} q_{\vec{k}} = 0,
\label{42-b}
\end{eqnarray}
where $\omega_{\tau, \vec{k}}$ is the dispersion relation for any mode $\vec{k}$, and is given by equation (\ref{omeeega}). In particular, for massless field we have
\begin{eqnarray}
\omega_{\tau, \vec{k}}^2\ =\ \sum_{i = 1}^{3} p_{i}^{2} k_{i}^{2}\ .
\label{disperClass}
\end{eqnarray}

Let us consider now a cosmological observer, whose 4-velocity is $u^{\mu} = (\sqrt{-1/g_{00}}, 0, 0, 0)$. This observer defines a local orthonormal frame $\{e^{\mu}_{a}\}$ as follows. First, the 4-velocity fixes the $0$th basis vector as $e^{\mu}_{0} = u^{\mu}$; then, we are free to orient the 3 spatial basis vectors as we wish - provided that they are all normalized and orthogonal to $e^{\mu}_{0}$. An obvious choice in our case is $e^{\mu}_{1} = (0, \sqrt{1/g_{11}}, 0, 0)$, $e^{\mu}_{2} = (0, 0, \sqrt{1/g_{22}}, 0)$ and $e^{\mu}_{3} = (0, 0, 0, \sqrt{1/g_{33}})$. We can write this basis compactly as
\begin{eqnarray}
e^{\mu}_{a} = \delta^{\mu}_{a} \frac{1}{\sqrt{|g_{aa}|}}\ ,
\label{orthobasis}
\end{eqnarray}
where there is no sum over $a$. Now, if $k_{\mu} = (\omega_{k}, k_{1}, k_{2}, k_{3})$ is the wave 4-vector of the quantum field, the 4-vector seen by the cosmological observer is given by its projection on the orthonormal basis:
\begin{eqnarray}
K_{a} = k_{\mu} e^{\mu}_{a} = \left(\dfrac{\omega_{k}}{\sqrt{-g_{00}}}, \dfrac{k_{1}}{\sqrt{g_{11}}}, \dfrac{k_{2}}{\sqrt{g_{22}}}, \dfrac{k_{3}}{\sqrt{g_{33}}}\right) .
\label{observedk}
\end{eqnarray}
The observed 3-velocity of the mode is then
\begin{eqnarray}
V^{i} = \frac{dK_{0}}{dK_{i}} = \sqrt{-\frac{g_{ii}}{g_{00}}} \frac{d \omega_{k}}{dk_{i}}\  ,
\label{observedv}
\end{eqnarray}
and the norm of this vector, i.e., $\|V\|^{2} := \eta_{ij} V^{i} V^{j}$, is simply
\begin{eqnarray}
\|V\|^{2}  = \sum_{i} (V^{i})^{2} = -\sum_{i} \frac{g_{ii}}{g_{00}} \left(\frac{d \omega_{k}}{dk_{i}}\right)^{2}.
\label{observeds}
\end{eqnarray}
This norm corresponds to the speed of the particle as measured by the cosmological observer. In the case at hand (i.e., for a massless scalar field on classical Bianchi I space-time), it is
\begin{eqnarray}
\|V\|^{2} = \sum_{i} \frac{1}{p_{i}^{2}} \left(\frac{k_{i} p_{i}^{2}}{\omega_{\tau, \vec{k}}}\right)^{2} = 1.
\label{flat}
\end{eqnarray}
As expected, the observed speed of massless particles coincides with the speed of light; this confirms the local Lorentz symmetry.

We can similarly study the case of a massless scalar field on the effective geometry. The corresponding wave equation (\ref{42-b}) for the effective geometry (\ref{effMetr}) can be further written as (see appendix \ref{ap-modes})
\begin{eqnarray}
\frac{d^{2}\mathcal{Q}_{\vec{k}}}{dT^{2}} + \Omega_{T,\vec{k}}^{2} \mathcal{Q}_{\vec{k}} = 0,
\label{WaveEq-Eff}
\end{eqnarray}
where $\mathcal{Q}_{\vec{k}}$, denotes the modified modes $q_{\vec{k}}$, and is defined in equation (\ref{ODE-1}) as
\begin{eqnarray}
\mathcal{Q}_{\vec{k}} := \frac{q_{\vec{k}}}{\sqrt{\langle \widehat{H}_{o}^{-1}\rangle}}\ .
\label{New-mode-Func}
\end{eqnarray}
Since equation (\ref{WaveEq-Eff}) has the form of a harmonic oscillator, the quantity $\Omega_{T,\vec{k}}(T)$ encodes the (modified) dispersion relation of the test field on the effective geometry (\ref{effMetr}). The modified dispersion relation $\Omega_{T,\vec{k}}(T)$ then is given by
\begin{eqnarray}
\Omega_{T,\vec{k}}^2(T)\ =\ \left(\bar{\omega}_{\vec{k}}^{2}-\frac{\beta^2}{4}-\frac{1}{2}\frac{d\beta}{dT} \right),
\label{disperEff}
\end{eqnarray}
where $\beta$ and $\bar{\omega}_{\vec{k}}$, are defined by equations (\ref{41-a}) and (\ref{41-b}), read,
\begin{align}
& \beta = -\frac{d\ln\langle \widehat{H}_{o}^{-1}\rangle}{dT} \ , \label{41a-Pheno} \\
& \bar{\omega}_{\vec{k}}^2 = \langle \widehat{H}_{o}^{-1}\rangle^2 \left(\sum_{i = 1}^{3} (\bar{p}_{i} k_{i})^{2} + |\bar{p}_{1} \bar{p}_{2} \bar{p}_{3}| m^{2}\right).
\label{41b-Pheno}
\end{align}
It should be noted that, in principle, $\Omega_{T,\vec{k}}$ in equation (\ref{disperEff}) can be imaginary, since in the $m = 0$ case one can make $\bar{\omega}_{\vec{k}}^{2}$ arbitrarily small by choosing small wave-vectors $\vec{k}$. However, two facts should be taken into account: first, the $\mathbb{T}^3$ topology of space introduces an IR cut-off; second, if the test field approximation holds, then $\widehat{H}_o^{-1}$ commutes with the total Hamiltonian, and can then be considered constant in $T$, so that $\beta$ vanishes\footnote{As for the $\vec{k} = 0$ mode of the massless field (for which one has $\Omega_{T, \vec{k}} = 0$), note that it is not an harmonic oscillator, but rather a spatially constant scalar field. We can think of it as the \emph{homogeneous} part of $\phi$, and use it instead of $T$ (indeed, the Klein-Gordon equation for $q_0$ reads $\partial^2 q_0/\partial T^2 = 0$, whose solution is linear in $T$).}.

Substituting equations (\ref{41a-Pheno}) and (\ref{41b-Pheno}) in equation (\ref{disperEff}), for a massless test field, we get
\begin{align}
\Omega_{T,\vec{k}}^2(T) & = \left[\langle \widehat{H}_{o}^{-1}\rangle\sum_i k_{i}^{2} \langle \widehat{H}_{o}^{-\frac{1}{2}} \widehat{p}_{i}^{2}(T) \widehat{H}_{o}^{-\frac{1}{2}} \rangle \right. \notag \\
& \left. - \frac{1}{4}\left(\frac{d\ln\langle \widehat{H}_{o}^{-1}\rangle}{dT}\right)^2+\frac{1}{2}\frac{d^2\ln\langle \widehat{H}_{o}^{-1}\rangle}{dT^2}\right].
\label{disperEff2}
\end{align}
As already pointed out, $\langle \widehat{H}_{0}^{-1} \rangle$ is independent of time $T$, and hence equation (\ref{disperEff2}) reduces to
\begin{align}
\Omega_{T,\vec{k}}^2(T) = \langle \widehat{H}_{o}^{-1}\rangle\sum_i k_{i}^{2}\langle \widehat{H}_{o}^{-\frac{1}{2}} \widehat{p}_{i}^{2}(T) \widehat{H}_{o}^{-\frac{1}{2}} \rangle .
\label{disperEff3}
\end{align}
Using equation (\ref{observeds}) we thus obtain the 3-velocity of modes propagating on the effective geometry as
\begin{align}
\|V\|^{2} & = -\sum_{i} \frac{\bar{g}_{ii}}{\bar{g}_{00}} \left(\frac{d \Omega_{T,\vec{k}}}{dk_{i}}\right)^{2}  \notag \\
& = \frac{1}{\Omega_{T,\vec{k}}^{2}} \sum_{i} k_{i}^{2} \langle \widehat{H}_{o}^{-1}\rangle \langle \widehat{H}_{o}^{-\frac{1}{2}} \widehat{p}_{i}^{2}(T) \widehat{H}_{o}^{-\frac{1}{2}} \rangle  \notag \\
& = 1 .
\label{effectives}
\end{align}
This equation confirms our expectation that, no Lorentz-violation is presented in our model herein.

%%%%%%%%%%%%%%%%%%%%%%%%%%%%%%%%%%%%%%%%%%%%%%%%%%%%%%%%%%%%%%%%%%%%%%%%%%%%%%%%%%%
\section{Conclusions and Discussion}
\label{conclusions}
%%%%%%%%%%%%%%%%%%%%%%%%%%%%%%%%%%%%%%%%%%%%%%%%%%%%%%%%%%%%%%%%%%%%%%%%%%%%%%%%%%%

In this work we considered an improved dynamics LQC of Bianchi I setting \cite{awe2} among the anisotropic class of models for a background cosmological quantum space-time, coupled with a massless scalar field. We developed the QFT of the test field on this background quantum geometry by considering a mode decomposition of such test field. It was shown that the QFT on this quantum space-time can be seen as a QFT on classical curved space-time; this background geometry is described by a classical effective metric, which can be thought of as emergent from the underlying quantum space-time once a mode of the field is coupled to it. It should be emphasized that this ``effective metric'' is of different nature than the usual semiclassical metric of LQC effective dynamics \cite{aps2, aps3}, since it arises from the peculiar procedure we deviced, involving the presence of a matter field. Despite high expectations, this effective geometry does not depend on any specific chosen mode of the test field, and therefore there is no Lorentz-violation. To investigate this in more details, we analyzed the dispersion relation for the matter field propagating on the background space-time; we have shown that it defines a speed \emph{in vacuo} which coincides with the speed of light.

A possible explanation for this could be that we have disregarded the back-reaction of the matter on the quantum geometry, i.e., we have been using the test field approximation (\ref{zerothstate}). It is possible to lift this approximation, by including the back-reaction via a Born-Oppenheimer (B-O) scheme. However, there are two caveats to this procedure:
\begin{itemize}
\item First, if the back-reaction of the mode on geometry is to be accounted, then one should, in principle, take into account the back-reaction of \emph{everything else} as well. Indeed, there is no reason to believe that, the back-reaction of a specific mode of the specific field has to be stronger than the others. Thus, by only considering this case, we are in essence restricting from QFT to quantum mechanics of a single harmonic oscillator. The physical relevance of this model can then be questioned.
\item In order  to apply the B-O scheme, the ``perturbation'' $\widehat{H}_{T, \vec{k}}$ in equation (\ref{53}) must act on the geometry sector at most as a multiplication. However, this is not the case since $\widehat{H}_{T, \vec{k}}$ involves $\widehat{H}_o$. To circumvent this problem, we might choose a different relational time: for the choice $\widetilde{T} = T/P_T$ the right hand side of the equation (\ref{53}) becomes $\widehat{H}_o^2/2 - \widehat{H}_{\tau, \vec{k}}$, whereas the ``perturbation'' $\widehat{H}_{\tau, \vec{k}}$ acts only by multiplication on the gravitational part. Now, this choice of time is uncommon in LQC, and someone might think that the result we got at the ``test field order'' does not hold for $\widetilde{T}$. In appendix \ref{app2} we will show that this is not the case: even by starting with $\widetilde{T}$, when disregarding the back-reaction, no Lorentz-violation is present.
\end{itemize}
We henceforth sketch the B-O procedure, bearing in mind these two remarks mentioned above, for the simplified FRW case (where $p_i = p=a^2$).

Following the ideas of B-O approximation, by denoting the gravitational degrees of freedom as ``heavy'' and the matter degrees of freedom as ``light'', one finds that the state of the system has the form
\begin{align} \label{cccA}
\Psi(\widetilde{T}, \lambda, q_{\vec{k}}) = \Psi_o(\widetilde{T}, \lambda) \otimes \psi(\widetilde{T}, q_{\vec{k}}) + \delta \Psi(\widetilde{T}, \lambda, q_{\vec{k}}) ,
\end{align}
where the correction is a non-simple tensor product of the two sectors:
\begin{align} \label{cccB}
\delta \Psi\ =\  \sum_{\alpha, i} f_{\alpha i} \varphi^o_\alpha \otimes \chi_i .
\end{align}
Here, $\varphi^o_\alpha$ is the geometry-eigenstate of $\Theta/2$ with eigenvalue $E^o_\alpha$, and $\chi_i$ is the matter-eigenstate of $\widehat{H}_{\tau, \vec{k}}$ with eigenvalue $\epsilon_i(p)$. The explicit form of the coefficients $f_{\alpha i}$ is given by
\begin{align} \label{cccC}
f_{\alpha i}\ :=\  \sum_{\beta \neq \alpha} c_\beta b_i \dfrac{\langle \varphi^o_{\alpha} | \epsilon_i(\widehat{p}) | \varphi^o_{\beta} \rangle}{E^o_\alpha - E^o_\beta}\ ,
\end{align}
where $c_\alpha$ and $b_i$ are the coefficients of the expansions of $\Psi_o$ on $\{\varphi^o_\alpha\}$, and of $\psi$ on $\{\chi_i\}$, respectively (so they are fixed by the $0$th order). Note that $\widehat{H}_{\tau, \vec{k}}$ acts on matter as the Hamiltonian of an harmonic oscillator with square frequency $k^2 p^2$, so its eigenvalues are linear in $k$, and thus $f_{\alpha i} \sim k$.

From equation (\ref{cccA}) it is hard to extract a dynamical equation for the matter only, since it is entangled with the geometry. Thus, we ``force'' the state to be again of the disentangled form: $\Psi = [\Psi_o + \delta \Psi_1] \otimes \psi$, where $\delta \Psi_1 = \sum_{\alpha, i} f_{\alpha i} \varphi^o_\alpha$. Plugging this new state into the constraint equation
\begin{align} \label{cccD}
-i \hbar \partial_{\widetilde{T}} \Psi = \left[\frac{1}{2} \Theta - \widehat{H}_{\tau, \vec{k}}\right] \Psi \ ,
\end{align}
and projecting on $\Psi_o$ (we may assume the correction $\delta \Psi_1$ to be orthogonal to $\Psi_o$), one finds a Shroedinger-like equation for $\psi$ only:
\begin{align}
i \hbar \partial_{\widetilde{T}} \psi = \frac{1}{2} \left[-\hbar^2 (1 + \sigma) \frac{d^2}{dq_{\vec{k}}^2} + k^2 \langle \widehat{p}^2 \rangle (1 + \sigma) q_{\vec{k}}^2\right] \psi .
\label{cccE}
\end{align}
Here, as before, $\langle \widehat{A} \rangle := \langle \Psi_o | \widehat{A}(\widetilde{T}) | \Psi_o \rangle$, and $\sigma$ is really an operator acting on matter; notice that, if we assume  $\psi$ to be (at least approximately) an eigenstate of $\sigma$, then we can think of $\sigma$ as a number\footnote{Formally, $\sigma$ is given by
\begin{align} \notag
\sigma = \langle \Psi_o | \widehat{H}_{\tau, \vec{k}} | \Psi_o \rangle^{-1} \langle \Psi_o | \widehat{H}_{\tau, \vec{k}} | \delta \Psi_1 \rangle.
\end{align}
}. Since $\sigma$ is proportional to $f_{\alpha i}$, its dependence on $k$ is linear, and thus we can write equation (\ref{cccE}) as
\begin{align}
i \hbar \partial_{\widetilde{T}} \psi = & \frac{1}{2} \left[-\hbar^2 (1 + \xi k) \frac{d^2}{dq_{\vec{k}}^2} + k^2 \langle \widehat{p}^2 \rangle (1 + \xi k) q_{\vec{k}}^2\right] \psi\ ,
\label{cccF}
\end{align}
where we introduced $\xi$ to make explicit the linear dependence of $\sigma$ on $k$. The number $\xi$ has dimensions of length (inverse of $k$), and is to be computed from $\sigma$ once the eigenequations for $\Theta$ and $\widehat{H}_{\tau, \vec{k}}$ have been solved.

Comparing equation (\ref{cccF}) with equation (\ref{44-c}) in the same way as done for the $0$th order (i.e., in the test field approximation case), one is led to an effective metric that indeed depends on $k$:
\begin{align} \label{cccG}
\bar{g}_{\mu \nu} dx^{\mu} dx^{\nu} = & -(1 + \xi k)^2 \langle \widehat{p}^2 \rangle^{3/2} d\widetilde{T}^2  \notag \\
& + \sqrt{\langle \widehat{p}^2 \rangle} \left(dx^2 + dy^2 + dz^2\right) .
\end{align}
Moreover, by carrying out the dispersion relation analysis for equation (\ref{cccF}) in the metric (\ref{cccG}), it is easy to check that the speed of the quanta with energy proportional to $k$ as measured by a cosmological observer is at $1$st order in $\xi$
\begin{align} \label{cccH}
\|V\| = 1 + \frac{1}{2} \xi k.
\end{align}
Thus, even though the above computation involved numerous approximations and assumptions, a prediction of Lorentz-violation is obtained, where the speed of particles deviates linearly from the standard value $c = 1$. Recent GRB observations \cite{OBS1} have probed linear deviations up to energies of order $10^7 J$, observing no Lorentz-violation. This value is just two orders of magnitude away from the plank scale ($E_{\text{Pl}} \approx 10^9 J$), at which quantum gravity is supposed to become important. This would correspond in our case to $\xi \approx \ell_{\text{Pl}}$. Hence, there is the concrete possibility that next generation of GRB observations be able to check our ``prediction''. Of course, if the prediction is disproved, this does not mean that LQG (or LQC) is wrong: we have used many approximations to get to equation (\ref{cccH}), and many of them could be not completely justified in the gravitational context. The result (\ref{cccH}) should thus be regarded as an indication, rather than a real prediction (especially in light of the fact that we are considering a quantum mechanical model, not the complete QFT, as explained in the first caveat above).

Another source of doubt comes from the following consideration: on one hand, we know that back-reaction of matter on classical geometry does not induce Lorentz-violation; on the other, from the very definition of the problem we have that $\xi$ accounts for both the classical and the quantum part of the back-reaction (formally, $\xi = \xi_C + \xi_Q$). Thus, it would be more appropriate if only $\xi_Q$ appeared in (\ref{cccH}). A possible way out of this probem lies in the observation that, while equation (\ref{cccF}) is obtained from the $1$st order in the back-reaction on quantum geometry, equation (\ref{44-c}) comes from an effective metric $\bar{g}_{\mu \nu}$ of the non-back-reacted Bianchi I type. Thus, the comparison of the two (which leads to (\ref{cccG}) and ultimately to (\ref{cccH})) is not completely justified. It would be more consistent to compare (\ref{cccF}) with the dynamical equation of an harmonic oscillator on an effective metric $\bar{g}^{(\vec{k})}_{\mu \nu}$ obtained as follows: take the harmonic ocillator (with $\vec{k}$-dependent frequency) on the FRW ``vacuum'' $g_{\mu \nu}$, and evaluate its energy-momentum tensor $\langle T_{\mu \nu} \rangle$; plug this into Einstein equation, and solve for the metric. This produces a classical metric $g^{(\vec{k})}_{\mu \nu}$ which ``knows'' about the matter (at $1$st order), i.e., it takes into account the $1$st order modification to classical geomtery due to the presence of the harmonic oscillator. One can then consider a generic effective metric $\bar{g}^{(\vec{k})}_{\mu \nu}$ of the kinematical family of $g^{(\vec{k})}_{\mu \nu}$, and find the dynamical equation for an harmonic oscillator on such an effective geometry. This is the equation to be compared with (\ref{cccF}) in order to single out the ``quantum part'' of the back-reaction, $\xi_Q$. Despite being an extremely complicate procedure, there is a change that the form of (\ref{cccH}) will not be altered much (possibly, only by the replacement of $\xi$ with $\xi_Q$), and in the worst case scenario a functionally more complicate $\vec{k}$-dependence should appear in $\|V\|$, so that Lorentz symmetry is in fact broken.

It should be noted that we never used the explicit form of $\widehat{H}_o$: it follows that this analysis can be applied to all quantum cosmologies which, absent matter, describe the state of geometry by a wave-function $\Psi_o$ evolving through $\widehat{H}_o$. Also, we would like to remark that this Lorentz deviation does \emph{not} come from the discrete nature of the quantum space-time: if we describe quantum gravity on a fixed graph, then it is not a surprise that Lorentz symmetry is broken, because from the point of view of matter propagation only certain modes are normal with respect to the given space-time ``crystal'' \cite{Tecotl}. On the contrary, we never invoked any feature of the Planck-scale space-time. Our only request is that it makes sense to consider only a limited number of gravitational degrees of freedom (in FRW case, just the scale factor). While this is of course a strong approximation, it does not involve any assertion on the microscopic discreteness of space-time.

Aside from this last indication of Lorentz-violation at $1$st order in back-reaction, we still have much to say. We proved that at $0$th order no symmetry-breaking is present in the Bianchi I case, thus extending the result of \cite{AKL}. Also, while the search for symmetry-breaking was our driving force, we exposed the first steps toward quantum field theory on quantum Bianchi I space-times. This is a result useful and interesting in its own right, and worth to be refined (in particular, more work should be done to be able to consider an infinite number of modes). It is to be said that developments in this direction are expected from the study of quantum perturbations \cite{AAN-pert}, which is an area of growing interest in the LQC community.

These are the first nontrivial findings about Lorentz-invariance in LQC, and a first step to understand it in full LQG. Previous attempts to study Lorentz-invariance in LQG have been done, but none of them attacked the problem from the indirect perspective we adopted: namely, via an effective ``classical'' metric that reproduces the same results as the semiclassical approximation of the underlying quantum theory of geometry.
We would like to focus the attention on the issue of Lorentz symmetry, which has not been clarified yet in the complete theory. Our result proves that no violation is present in Bianchi I case (at least when back-reaction can be disregarded), and a linear deviation could be present once the back-reaction is taken into account. However, such findings do not solve the problem. In this search, the tools of ``effective metric'' and dispersion relation analysis can be used for more realistic cases of matter coupling (such as the electromagnetic field).
We hope this work will trigger the question of Lorentz-invariance in LQG, as this approach to quantum gravity does not yet give a
definitive statement. Such a statement could be in principle used to falsify the theory, especially considering the great effort on the observational side \cite{OBS1,ACamelia1,ACamelia2,ACamelia3}. We would like to know where LQG stands in this context. Of course more work is needed, and we hope to start a discussion in this direction within the LQG/LQC community.

%%%%%%%%%%%%%%%%%%%%%%%%%%%%%%%%%%%%%%%%%%%%%%%%
\section{Acknowledgments}
%%%%%%%%%%%%%%%%%%%%%%%%%%%%%%%%%%%%%%%%%%%%%%%%

We are grateful to Kristina Giesel for the enlightening discussion about Born-Oppenheimer approximation, and to Abhay Ashtekar for his comments about its applicability. The authors would also like to thank the unknown referee for useful remarks and suggestions. This work was partially supported by the grant 182/N-QGG/2008/0 (PMN) of Polish Ministerstwo Nauki i Szkolnictwa Wy\.{z}szego. YT thanks the ESF sponsored network `Quantum Geometry and Quantum Gravity' for a short visit grant to collaborate with the Warsaw group. He is supported by the Portuguese Agency Funda\c{c}\~{a}o para a Ci\^{e}ncia e Tecnologia through the fellowship SFRH/BD/43709/2008. He also would like to thank the warm hospitality of the department of theoretical physics of the University of Warsaw, where the idea of this work was initiated.

%%%%%%%%%%%%%%%%%%%%%%%%%%%%%%%%%%%%%%%%%%%%%%%%%%%%%%%%%%%%%%%%
\appendix
%%%%%%%%%%%%%%%%%%%%%%%%%%%%%%%%%%%%%%%%%%%%%%%%%%%%%%%%%%%%%%%%

\section{Mode Decomposition}
\label{ap-modes}

Let us consider a decomposition of $\phi_{\vec{k}}$ and $\pi_{\vec{k}}$ as
\begin{eqnarray}
\phi_{\vec{k}}\ &=\ \frac{1}{\sqrt{2}}\left(\phi_{\vec{k}}^{(1)} + i \phi_{\vec{k}}^{(2)}\right), \notag \\
\pi_{\vec{k}}\ &=\ \frac{1}{\sqrt{2}}\left(\pi_{\vec{k}}^{(1)} + i \pi_{\vec{k}}^{(2)}\right).
\label{31}
\end{eqnarray}
Since the reality conditions are satisfied, not all variables in (\ref{31}) are independent. In particular, we have
\begin{align}
\phi_{-\vec{k}}^{(1)} = \phi_{\vec{k}}^{(1)}, \ \ \ \ \ \phi_{-\vec{k}}^{(2)} = -\phi_{\vec{k}}^{(2)}\ , \notag \\
\pi_{-\vec{k}}^{(1)} = \pi_{\vec{k}}^{(1)}, \ \ \ \ \ \pi_{-\vec{k}}^{(2)} = -\pi_{\vec{k}}^{(2)}\ .
\label{32}
\end{align}
Because there exist relations between the ``positive" and ``negative" modes $\vec{k}$ and $-\vec{k}$, one can split the lattice $\mathcal{L}$ into positive and negative parts:
\begin{align}
\mathcal{L}_{+}\ & =\ \{\vec{k} : k_{3} > 0\} \cup \{\vec{k} : k_{3} = 0, k_{2} > 0\}  \notag \\
&\ \ \ \ \  \cup \{\vec{k} : k_{3} = k_{2} = 0, k_{1} > 0\}, \notag \\
\mathcal{L}_{-}\ &=\ \{\vec{k} : k_{3} < 0\} \cup \{\vec{k} : k_{3} = 0, k_{2} < 0\}  \notag \\
&\ \  \ \ \ \cup \{\vec{k} : k_{3} = k_{2} = 0, k_{1} < 0\} .
\label{33-R}
\end{align}
Observe that, if $\vec{k} \in \mathcal{L}_{+}$, then $-\vec{k} \in \mathcal{L}_{-}$. Using this fact, we are now able to split the summation in the Hamiltonian (\ref{30}) into its positive and negative parts as $H_{\phi} = H_{\phi}^{+} + H_{\phi}^{-}$.
In $H_{\phi}^{-}$, we change the dummy index $\vec{k}$ into $-\vec{k}$, so that we can convert its sum from $\sum_{-\vec{k} \in \mathcal{L}_{-}}$ to $\sum_{\vec{k} \in \mathcal{L}_{+}}$, the same as the sum in $H_{\phi}^{+}$. Thus, we get
\begin{align} %\label{34-b}
& H_{\phi}(x_0)  = \frac{1}{2} \sum_{\vec{k} \in \mathcal{L}_{+}} \frac{N_{x_{0}}}{\sqrt{|p_{1} p_{2} p_{3}|}} \left[\overline{\pi_{\vec{k}}} \pi_{\vec{k}} + \overline{\pi_{-\vec{k}}} \pi_{-\vec{k}} \right. \notag \\
& \ \ \ \  + \left. \left(\sum_{i = 1}^{3} (p_{i} k_{i})^{2} + |p_{1} p_{2} p_{3}| m^{2}\right) \left(\overline{\phi_{\vec{k}}} \phi_{\vec{k}} + \overline{\phi_{-\vec{k}}} \phi_{-\vec{k}}\right)\right] \notag \\
& \ \ \  =  \frac{1}{4} \sum_{\vec{k} \in \mathcal{L}_{+}} \sum_{\sigma = 1, 2} \frac{N_{x_{0}}}{\sqrt{|p_{1} p_{2} p_{3}|}} \left[(\pi_{\vec{k}}^{(\sigma)})^{2} + (\pi_{-\vec{k}}^{(\sigma)})^{2} \right.
\notag \\
%&  [(\pi_{\vec{k}}^{(\sigma)})^{2} + (\pi_{-\vec{k}}^{(\sigma)})^{2} + \notag \\
&\ \ \ \ + \left. \left(\sum_{i = 1}^{3} (p_{i} k_{i})^{2} + |p_{1} p_{2} p_{3}| m^{2}\right) \left((\phi_{\vec{k}}^{(\sigma)})^{2} + (\phi_{-\vec{k}}^{(\sigma)})^{2}\right)\right] \notag \\
& \ \ \  =  \frac{1}{2} \sum_{\vec{k} \in \mathcal{L}_{+}} \sum_{\sigma = 1, 2} \frac{N_{x_{0}}}{\sqrt{|p_{1} p_{2} p_{3}|}} \left[(\pi_{\vec{k}}^{(\sigma)})^{2} \right.  \notag \\
& \ \ \  \ \ \ \ \left.  + \left(\sum_{i = 1}^{3} (p_{i} k_{i})^{2} + |p_{1} p_{2} p_{3}| m^{2}\right) (\phi_{\vec{k}}^{(\sigma)})^{2}\right],
\label{34}
\end{align}
in which we have used equation (\ref{31}) in the second step, and equation (\ref{32}) in the third step. In the final expression all the modes are independent, so this is really a collection of independent harmonic oscillators. We can do even better if we think of $\sigma$ as the ``sign'' of $\vec{k}$. In this way we can combine the sums $\sum_{\vec{k} \in \mathcal{L}_{+}}$ and $\sum_{\sigma = 1, 2}$ into a single one: $\sum_{\vec{k} \in \mathcal{L}}$.
In terms of these variables, the Hamiltonian (\ref{34}) has the form
\begin{eqnarray} \label{37}
H_{\phi}(x_{0})\ =\ \sum_{\vec{k} \in \mathcal{L}} H_{\vec{k}}(x_{0}),
\end{eqnarray}
where $H_{\vec{k}}(x_0)$ is given by
\begin{align}
H_{\vec{k}}\ & =\  \frac{N_{x_{0}}}{2 \sqrt{|p_{1} p_{2} p_{3}|}}  \notag \\
& \times \left[p_{\vec{k}}^{2} + \left(\sum_{i = 1}^{3} (p_{i} k_{i})^{2} + |p_{1} p_{2} p_{3}| m^{2}\right) q_{\vec{k}}^{2}\right],
\label{H-k}
\end{align}
and we defined the following real variables:
\begin{eqnarray}
q_{\vec{k}} = \left\{
\begin{array}{ll}
\phi_{\vec{k}}^{(1)}		&		\mbox{if} \ \vec{k} \in \mathcal{L}_{+}
\\
\\
\phi_{-\vec{k}}^{(2)}		&		\mbox{if} \ \vec{k} \in \mathcal{L}_{-}
\end{array}
\right., \ \ \ \ p_{\vec{k}} = \left\{
\begin{array}{ll}
\pi_{\vec{k}}^{(1)}		&		\mbox{if} \ \vec{k} \in \mathcal{L}_{+}
\\
\\
\pi_{-\vec{k}}^{(2)}		&		\mbox{if} \ \vec{k} \in \mathcal{L}_{-}
\end{array}
\right.
\label{35}
\end{eqnarray}
Using the relation $\{\phi_{\vec{k}}, \pi_{\vec{k}'}\} = \delta_{\vec{k}, -\vec{k}'}$ together with equation (\ref{31}), it then follows
\begin{eqnarray}
\{\phi_{\vec{k}}^{(1)}, \pi_{-\vec{k}'}^{(1)}\} = \delta_{\vec{k} \vec{k}'}, \ \ \ \ \{\phi_{\vec{k}}^{(2)}, \pi_{-\vec{k}'}^{(2)}\} = -\delta_{\vec{k} \vec{k}'} .
\end{eqnarray}
So that, we can compute $\{q_{\vec{k}}, \pi_{\vec{k}'}\}$ as follows: For all $\vec{k}, \vec{k}' \in \mathcal{L}_{+}$ we get,
\begin{align}
\{q_{\vec{k}}, \pi_{\vec{k}'}\} = \{\phi_{\vec{k}}^{(1)}, \pi_{\vec{k}'}^{(1)}\} = \{\phi_{\vec{k}}^{(1)}, \pi_{-\vec{k}'}^{(1)}\} = \delta_{\vec{k} \vec{k}'}\ ;
\end{align}
for all $\vec{k} \in \mathcal{L}_{+}$, and $\vec{k}' \in \mathcal{L}_{-}$ we have
\begin{align}
\{q_{\vec{k}}, \pi_{\vec{k}'}\} = \{\phi_{\vec{k}}^{(1)}, \pi_{-\vec{k}'}^{(2)}\} = 0\ ;
\end{align}
for all $\vec{k} \in \mathcal{L}_{-}$, and $\vec{k}' \in \mathcal{L}_{+}$ we have,
\begin{align}
\{q_{\vec{k}}, \pi_{\vec{k}'}\} = \{\phi_{-\vec{k}}^{(2)}, \pi_{\vec{k}'}^{(1)}\} = 0 \ ;
\end{align}
and finally, for all $\vec{k}, \vec{k}' \in \mathcal{L}_{-}$ we get
\begin{align}
\{q_{\vec{k}}, \pi_{\vec{k}'}\} = \{\phi_{-\vec{k}}^{(2)}, \pi_{-\vec{k}'}^{(2)}\} = -\{\phi_{\vec{k}}^{(2)}, \pi_{-\vec{k}'}^{(2)}\} = \delta_{\vec{k} \vec{k}'}.
\end{align}
We then conclude that $q_{\vec{k}}$ and $p_{\vec{k}}$ are canonically conjugates:
\begin{eqnarray}
\{q_{\vec{k}}, p_{\vec{k}'}\}\ =\  \delta_{\vec{k} \vec{k}'}.
\label{36}
\end{eqnarray}
Therefore, each pair $(q_{\vec{k}}, p_{\vec{k}})$ describes a harmonic oscillator, decoupled from all others and evolving according to $H_{\vec{k}}(x_{0})$ defined in equation (\ref{H-k}).

From this knowledge, we can go back and reconstruct the equation of motion for $q_{\vec{k}}$. From Hamilton equations
\begin{align} \label{38}
\dfrac{dq_{\vec{k}}}{dx_{0}} & = \{q_{\vec{k}}, H_{\vec{k}}\} = \dfrac{N_{x_{0}}}{\sqrt{|p_{1} p_{2} p_{3}|}} p_{\vec{k}} \notag \\
\dfrac{dp_{\vec{k}}}{dx_{0}} & = \{p_{\vec{k}}, H_{\vec{k}}\} = -\dfrac{N_{x_{0}}}{\sqrt{|p_{1} p_{2} p_{3}|}} \notag \\
& \ \ \ \times \left(\sum_{i = 1}^{3} (p_{i} k_{i})^{2} + |p_{1} p_{2} p_{3}| m^{2}\right) q_{\vec{k}},
\end{align}
one gets
\begin{align} \label{39}
\frac{d^{2}q_{\vec{k}}}{dx_{0}^{2}} & =  \frac{d}{dx_{0}} \left(\dfrac{N_{x_{0}}}{\sqrt{|p_{1} p_{2} p_{3}|}}\right) p_{\vec{k}} + \dfrac{N_{x_{0}}}{\sqrt{|p_{1} p_{2} p_{3}|}} \frac{dp_{\vec{k}}}{dx_{0}} \notag \\
& =  \frac{d}{dx_{0}} \ln\left(\dfrac{N_{x_{0}}}{\sqrt{|p_{1} p_{2} p_{3}|}}\right) \frac{dq_{\vec{k}}}{dx_{0}} - \dfrac{N^{2}_{x_{0}}}{|p_{1} p_{2} p_{3}|}  \notag \\
&  \  \ \ \ \times \left(\sum_{i = 1}^{3} (p_{i} k_{i})^{2} + |p_{1} p_{2} p_{3}| m^{2}\right) q_{\vec{k}}.
\end{align}
For simplicity, we rewrite equation (\ref{39}) as
\begin{eqnarray}
\frac{d^{2}q_{\vec{k}}}{dx_{0}^{2}} + \beta \frac{dq_{\vec{k}}}{dx_{0}} + \omega_{\vec{k}}^{2} q_{\vec{k}} = 0,
\label{40}
\end{eqnarray}
where the time-dependent parameters $\beta(x_0)$ and $ \omega_{\vec{k}}(x_0)$ are given by
\begin{align}
\beta\ &=\  -\frac{d}{dx_{0}} \ln\left(\dfrac{N_{x_{0}}}{\sqrt{|p_{1} p_{2} p_{3}|}}\right), \label{41-a} \\
\omega_{\vec{k}}^2\ &=\ \dfrac{N^{2}_{x_{0}}}{|p_{1} p_{2} p_{3}|} \left(\sum_{i = 1}^{3} (p_{i} k_{i})^{2} + |p_{1} p_{2} p_{3}| m^{2}\right).
\label{41-b}
\end{align}

In order to get the dispersion relation for mode $\vec{k}$, it is better to write the equation (\ref{40}) in its \emph{normal form}. To do this, let us consider a new variable $\mathcal{Q}_{\vec{k}}$:
\begin{equation}
\mathcal{Q}_{\vec{k}}:=q_{\vec{k}}\exp\left(\frac{1}{2}\int^{x_0} \beta(x_0')dx_0'\right).
\label{ODE-1}
\end{equation}
With this new variable, equation (\ref{40}) can be rewritten as
\begin{eqnarray}
\frac{d^{2}\mathcal{Q}_{\vec{k}}}{dx_{0}^{2}} +  \Omega_{\vec{k}}^{2} \mathcal{Q}_{\vec{k}} = 0,
\label{40-simplified}
\end{eqnarray}
where $\Omega_{\vec{k}}^{2}$ is given by
\begin{eqnarray}
\Omega_{\vec{k}}^{2}=\left(\omega_{\vec{k}}^{2}-\frac{\beta^2}{4}-\frac{1}{2}\frac{d\beta}{dx_0}\right).
\label{40-simplified2}
\end{eqnarray}

In the case of harmonic time $x_{0} = \tau$, $\beta = 0$ (since $N_{x_{0}} = \sqrt{|p_{1} p_{2} p_{3}|}$), and equation (\ref{40-simplified}) reduces to
\begin{eqnarray}
\frac{d^{2}q_{\vec{k}}}{d\tau^{2}} + \omega_{\tau, \vec{k}}^{2} q_{\vec{k}} = 0,
\label{42}
\end{eqnarray}
with
\begin{eqnarray}
\omega_{\tau, \vec{k}}^2 = \sum_{i = 1}^{3} (p_{i} k_{i})^{2} + |p_{1} p_{2} p_{3}| m^{2}.
\label{disperClass-b}
\end{eqnarray}

%%%%%%%%%%%%%%%%%%%%%%%%%%%%%%%%%%%%%%%%%%%%%%%%%%%%%%%%%%%%%%%%%%%%%%%%
\section{A different Choice of Time}
\label{app2}
%%%%%%%%%%%%%%%%%%%%%%%%%%%%%%%%%%%%%%%%%%%%%%%%%%%%%%%%%%%%%%%%%%%%%%%%

It could be argued that the main result of our work (namely, the independence of the effective metric on the field mode) is due to the approximations was taken herein (cf. section \ref{BI-Effective}). For this reason, we here summarize such approximations:
\begin{itemize}
\item The expansion of the square root in the equation (\ref{50}) to get equation (\ref{53}).
\item The test field approximation $\Psi(T, \vec{\lambda}, q_{\vec{k}}) = \Psi_{o}(T, \vec{\lambda}) \otimes \psi(T, q_{\vec{k}})$, that is, the assumption that back-reaction can be disregarded.
\end{itemize}
As discussed in section \ref{conclusions}, a possible way to lift the second approximation is to make use of B-O scheme. However, to do so one needs to choose a different relational time. If this can seem awkward at first glance, one should remember that time in general relativity is simply a gauge parameter. Moreover, with the choice
\begin{align}
\widetilde{T}\ :=\  \frac{T}{P_{T}}\ ,
\end{align}
we find again the same result as above, thus making it more robust. Incidentally, this choice also lifts the first approximation, that is, the square root in equation (\ref{50}) disappears.

The new relational time parameter $\widetilde{T}$ is again a good parameter for the gauge flow of $C$. Indeed, the conjugate momentum of $\widetilde{T}$ is $P_{\widetilde{T}} = P_{T}^{2}/2$, so we have
\begin{align}
C_{T} = \frac{P_{T}^{2}}{2} = P_{\widetilde{T}},
\end{align}
and hence $d\widetilde{T}/d\tau = 1$, and $dP_{\widetilde{T}}/d\tau = 0$, from which it follows that in fact
\begin{align}
\widetilde{T} = \tau .
\end{align}
The lapse function of $\widetilde{T}$ is then
\begin{align}
N_{\widetilde{T}} = N_{\tau} = \sqrt{|p_{1} p_{2} p_{3}|},
\end{align}
so that, instead of equation (\ref{20}) we have
\begin{align}
g_{\mu \nu} dx^{\mu} dx^{\nu} = |p_{1} p_{2} p_{3}| \left(-d\widetilde{T}^{2} + \sum_{i = 1}^{3} \frac{(dx^{i})^{2}}{p_{i}^{2}}\right).
\end{align}
Quantization of the geometrical part is then on the lines of the $T$-case: one promotes $\widetilde{T}$ and $P_{\widetilde{T}}$ to operators, but this time in $C_{\text{geo}}$ only a first derivative in $\widetilde{T}$ appears. Physical states are thus those $\Psi_{o}(\widetilde{T}, \vec{\lambda}) \in \mathcal{H}_{T} \otimes \mathcal{H}_{\text{gr}}$ that verify the Schroedinger-like equation
\begin{align}
-i\hbar \partial_{\widetilde{T}} \Psi_{o}(\widetilde{T}, \vec{\lambda}) = \frac{1}{2} \Theta \Psi_{o}(\widetilde{T}, \vec{\lambda}).
\end{align}
For the QFT on quantum geometry, we consider the full Hamiltonian constraint for mode $\vec{k}$ of the matter field $\phi$ and harmonic time $\tau$:
\begin{align}
\widehat{C}_{\tau, \vec{k}} = \widehat{C}_{\text{geo}} \otimes \mathbb{I}_{\vec{k}} + \mathbb{I}_{\widetilde{T}} \otimes \widehat{H}_{\tau, \vec{k}},
\end{align}
where $\widehat{H}_{\tau, \vec{k}}$ acts on both gravitational and matter degrees of freedom, and is explicitly given by (for massless scalar field)
\begin{align}
\widehat{H}_{\tau, \vec{k}} = \frac{1}{2} \left[\widehat{p}_{\vec{k}}^{2} + \left(\sum_{i = 1}^{3} \widehat{p}_{i}^{2} k_{i}^{2}\right) \widehat{q}_{\vec{k}}^{2}\right].
\end{align}
Physical states are hence those $\Psi(\widetilde{T}, \vec{\lambda}, q_{\vec{k}})$ that solve the Schroedinger-like equation
\begin{align}
-i\hbar \partial_{\widetilde{T}} \Psi(\widetilde{T}, \vec{\lambda}, q_{\vec{k}}) & = \frac{1}{2} \left[\Theta -2 \widehat{H}_{\tau, \vec{k}}\right] \Psi(\widetilde{T}, \vec{\lambda}, q_{\vec{k}})  \notag \\
& = \frac{1}{2} \left[\widehat{H}_{0}^{2} -2 \widehat{H}_{\tau, \vec{k}}\right] \Psi(\widetilde{T}, \vec{\lambda}, q_{\vec{k}}).
\end{align}
Note the difference with the $T$-case: here we do not have any square root on the r.h.s. Since classically $H_{\tau, \vec{k}} = H_{\widetilde{T}, \vec{k}}$, one can write $\widehat{H}_{\tau, \vec{k}} = \widehat{H}_{\widetilde{T}, \vec{k}}$. Approximating again the state as the disentangled tensor product of geometry and matter (where now $\Psi_o$ obeys $-i \hbar \partial_{\widetilde{T}} \Psi_o = \frac{1}{2} \widehat{H}_o^2 \Psi_o$), and projecting on $\Psi_o$, we find a Schroedinger-like equation for $\psi(T, q_{\vec{k}})$:
\begin{align}
i\hbar \partial_{\widetilde{T}} \psi = \frac{1}{2} \left[-\hbar^{2} \frac{\partial^{2}}{\partial q_{\vec{k}}^{2}} + \left(\sum_{i = 1}^{3} \langle \widehat{p}_{i}^{2}(\widetilde{T}) \rangle k_{i}^{2}\right) q_{\vec{k}}^{2}\right] \psi .
\end{align}
By the comparison with the equation obtained for $\psi$ in the case of an effective classical geometry $\bar{g}_{\mu \nu}$, (\ref{44-c}), we read off the effective terms:
\begin{align}
\frac{\bar{N}_{\widetilde{T}}}{\sqrt{|\bar{p}_{1} \bar{p}_{2} \bar{p}_{3}|}} = 1, \ \ \ \bar{p}_{i}^{2} = \langle \widehat{p}_{i}^{2}(\widetilde{T})\rangle .
\end{align}
Therefore, the effective geometry probed by the mode $\vec{k}$ of the field $\phi$ is obtained as
\begin{align}
\bar{g}_{\mu \nu} dx^{\mu} dx^{\nu} & = \sqrt{|\langle \widehat{p}_{1}^{2}(\widetilde{T}) \rangle \langle \widehat{p}_{2}^{2}(\widetilde{T}) \rangle \langle \widehat{p}_{3}^{2}(\widetilde{T}) \rangle|} \notag \\
& \ \ \ \ \ \times \left[-d\widetilde{T}^{2} + \sum_{i = 1}^{3} \frac{(dx^{i})^{2}}{\langle \widehat{p}_{i}^{2}(\widetilde{T}) \rangle}\right].
\end{align}
Once again, $\bar{g}_{\mu \nu}$ does not depend on the mode $\vec{k}$: even with this choice of time, which allows to lift the square root approximation and to push the analysis to the inclusion of back-reaction, we see no violation of Lorentz symmetry at the test field order.

%%%%%%%%%%%%%%%%%%%%%%%%%%%%%%%%%%%%%%%%%%%%%%%%%%%%%%%%%%%%%%%%%%%%%%%%%%%%%%%%%%%%%%%%%

\end{document}